\documentclass[12pt,twoside]{article}
\usepackage{fleqn,espcrc1}

\newcommand{\AmS}{{\protect\the\textfont2
  A\kern-.1667em\lower.5ex\hbox{M}\kern-.125emS}}
\def\vec#1{\mathbf{#1}}  

\title{Few-Body Problems in Hadron Spectroscopy
\footnote{Invited Talk at the 17th European Conference on Few-Body 
Physics, 11--16 September 2000, Evora, Portugal, to appear in Nuclear 
Physics A}
}

\author{Jean-Marc Richard%
\address{Institut des Sciences Nucl\'eaires, 
        Universit\'e Joseph Fourier--CNRS-IN2P3, \\ 
        53, avenue des Martyrs, 38026 Grenoble Cedex, France}}
       
\begin{document}
\maketitle
\begin{abstract}
Some rigorous results can be derived using a very simple approach to
hadron spectroscopy, in which a static potential is associated with
non-relativistic kinematics.  Several regularities of the experimental
spectrum are explained by such models.  It is underlined that certain
methods developed for hadronic physics have found applications in
other fields, in particular atomic physics.  A few results can be
extended to cases involving spin-dependent forces or relativistic
kinematics.
\end{abstract}
\section{Introduction}
\label{se:intro}
The dynamics of light quarks is presumably very intricate, with
non-pertur\-bative effects even at short distances, and
highly-relativistic motion of the constituents inside hadrons. 
Nevertheless, it is interesting to consider a fictitious world, where
the quark dynamics is governed by a simple Hamiltonian with
non-relativistic kinematics supplemented by a static,
flavour-independent potential.  Most regularities derived from the
properties of the Schr\"odinger equation are observed in the
experimental spectrum.  This suggests that the QCD theory of quark
confinement should lead to similar regularities.

Several results derived in the context of quark models have been
successfully applied to other few-body problems, for instance in
atomic physics.  Examples will be given in this review.

A challenge consists of extending the theorems on level order, convexity, 
etc., to less naive Hamiltonians with spin-dependent forces and 
relativistic kinematics. Some of the first results will be mentioned.

\section{Results on mesons}
\label{se:mesons}
Studies of the quark model have been stimulated by the discovery of
$J\!/\!\Psi$ and $\Upsilon$ resonances and their excitations and, in
particular, the simultaneous description of both spectra using the
{\sl same} potential (flavour independence).  Many rigorous results
have been summarized in the reviews by Quigg and Rosner \cite{QR1} and
by Martin and Grosse \cite{GM1,GMbook}.  A few examples are given
in this section, dealing with energy levels.  A result on the wave function will
be mentioned in Sec.~\ref{se:short}.

Current potentials reproduce the observed pattern  that
$E(1P)<E(2S)$.  Within the notation ($n,\ell)$ adopted here for 
quarkonium, the
radial wave function has $n-1$ nodes and the principal quantum number
of atomic physics is $n+\ell$.  It has been proved that
$E(n+1,\ell) > E(n,\ell+1)$ if $\Delta V>0$, and the reverse if $\Delta
V<0$.  The Coulomb degeneracy is recovered as a limiting case.  From
the Gauss theorem, the sign of the Laplacian $\Delta V=r^{-2}[r^2V']'$
reflects whether the charge $Q(r)$ seen at distance $r$ grows
(asymptotic freedom), decreases or remains constant.

This ``Coulomb theorem'' can be applied successfully to muonic atoms,
which are sensitive to the size of the nucleus ($Q(r)\nearrow$), and
to alkaline atoms whose last electron penetrates the inner electron
shells ($Q(r)\searrow$). The latter property also hold for the 
metastable $\bar{p}-{\rm He}$ molecules described by Korobov at this 
conference \cite{Korobov}.

Another theorem explains how the harmonic oscillator (h.o.) 
degeneracy $E(n+1,\ell)=E(n,\ell+2)$ is broken. A strict inequality 
is obtained if the sign of $V''$ is constant \cite{GMbook}.

In both the complete Hamiltonian $\vec{p}_{1}^2/(2
m_{1})+\vec{p}_{2}^2/(2 m_{2})+V(r_{12})$ or its reduced version
$\vec{p}^2/(2\mu)+V(r)$, the individual inverse masses, $m_{i}^{-1}$, or
the inverse reduced mass, $\mu^{-1}$, enter linearly through a positive operator
$\vec{p}^2$.  This implies that each energy level is an increasing
function of this inverse mass, and that the ground-state energy (or
the sum of first levels) is a concave function of this variable
\cite{Thir}.  Let us mention two among the many applications.

The dynamics of heavy quarkonia $Q\overline{Q}$ is very sensitive to
the change from $Q=c$ to $Q=b$, while open-flavour mesons are
dominated by the light quark mass.  Thus $B-D$ is essentially the
difference between quark masses $b-c$, while the effect of this
difference is reduced in $\Upsilon-\Psi$ by the change in the binding
energy.  After removing or estimating the hyperfine effects, one gets
from the data \cite{PDG}
\begin{equation}
    {1\over2}(\Upsilon-\Psi)\simeq 3.2\;{\rm GeV} < (B-D)\simeq 3.3\; 
    {\rm GeV}.
 \end{equation}
The second example is concerned with comparing symmetric mesons
$Q\overline{Q}$ with hidden flavour to asymmetric mesons
$Q'\overline{Q}$ with open flavours.  As $(b\bar{c})$, for instance,
has an inverse reduced mass which is the average between that of
$(b\bar{b})$ and that of $(c\bar{c})$, then \cite{BeMa,Massdependence}
$(b\bar{c})>(b\bar{b}+c\bar{c})/2$ is expected.  Moreover, ${\rm
d}E/{\rm d}\mu^{-1}=\langle\vec{p}^2\rangle/2$, and this latter matrix
element can be estimated from the excitation spectrum and bounded
along its path from $(b\bar{b})$ or $(c\bar{c})$ to $(b\bar{c})$. One
eventually ends with a rigorous window for the mass of the
spin-averaged ground state\cite{GMbook,Bagan}
\begin{equation}
    \label{bcbar1}
    6.26<(b\bar{c})<6.43\;{\rm GeV}.
\end{equation}
Of course, hyperfine effects should be added in 
more precise tests. The accuracy of the 
first measurement \cite{PDG}
\begin{equation}
    \label{bcbar3}
    (b\bar{c})=6.4\pm0.4\;{\rm GeV}
\end{equation}
is not yet sufficient to test flavour independence.

Convexity in the inverse reduced mass also implies that 
$(b\bar{q}_{2})+(c\bar{q}_{1})<(\bar{q}_{1})+(c\bar{q}_{2})$
if $b>c$ and $q_{2}>q_{1}$. In particular,
\begin{equation}
    \label{bcbar2}
   (b\bar{c})> (b\bar{s})+(c\bar{c})-(c\bar{s})\simeq 6.52\;{\rm GeV},
\end{equation}
which is less efficient but much more easily obtained than the r.h.s.\ of 
Eq.~(\ref{bcbar1}).

Note that an inequality such as $\mathrm{D}_{2}+\mathrm{H}_{2}\le 2
\mathrm{HD}$ should hold for variants of the hydrogen molecule
involving isotopes, as they experience the same Born--Oppenheimer
potential.

One of the ``sides'' of the Coulomb theorem can be adapted to the 
Klein--Gordon equation \cite{GMbook}. If $\Delta V\le 0$, then 
$E(n+1,\ell)<E(n,\ell+1)$.
\section{Level order of baryon spectra}
\label{se:baryons:order}
For many years, the only widespread knowledge of the 3-body problem 
was the harmonic oscillator. This remains true outside the few-body 
 community. The discussion on baryon excitations is thus often restricted to 
situations where $V=\sum v(r_{ij})$, with $v(r)=Kr^2+\delta 
v$, and $\delta v$ treated as a correction. 

First-order perturbation theory is usually excellent, especially if
the oscillator strength $K$ is variationally adjusted to minimise the
magnitude of the corrections.  However, when first-order perturbation
is shown (or claimed) to produce a crossing of levels, one is
reasonably worried about higher-order terms, and a more rigorous
treatment of the energy spectrum becomes desirable.

In this context, it should be recalled that in the $N$ and $\Delta$
sector (not in the $\Lambda$ one!), the radial excitation with
positive parity (Roper resonance) comes first, and the orbital
excitation with positive parity just above.  On the other hand, any
simple potential such as $v(r)=\lambda r-a/r$ gives the opposite
ordering.  This problem has been known for many years.  It has been
vigorously revisited by Glozman and collaborators
\cite{Glozman:2000vd} and many others.  The true nature of light-quark
dynamics is outside the scope of this review, which will be focused on
the level order in 3-body potential models.

A  decomposition better than $V=\sum K r_{ij}^2+\delta v$ is provided 
by the generalised partial-wave expansion 
\begin{equation}
    \label{pot-hyper}
    V=V_{0}(\rho)+\delta V,
\end{equation}
where $\rho\propto(r_{12}^2+r_{23}^2+r_{31}^2)^{1/2}$ is the 
hyperradius. The last term $\delta V$ gives a very small correction 
to the first levels \cite{JMRrep}. With the hyperscalar potential 
$V_{0}$ only, the wave function reads  $\Psi=\rho^{-5/2} u(\rho) 
P_{[L]}(\Omega)$, where the last factor contains the ``grand-angular'' 
part. The energy and the hyperradial part are governed by
\begin{equation}
    \label{radial-hyper}
    u''(\rho)-{\ell(\ell+1)\over \rho^2}u(\rho)
    +m[E-V_{0}(\rho)] u(\rho)=0,
\end{equation}
very similar to the usual radial equations of the 2-body problem,
except that the effective angular momentum is now $\ell=3/2$ for the
ground-state and its radial excitations and $\ell=5/2$ for the first
orbital excitation with negative parity.  The Coulomb theorem holds
for non-integer $\ell$.  If $\Delta V>0$, then $E(2S)>E(1P)$, i.e.,
the Roper comes above the orbital excitation \cite{HogRic}.  Note that
a three-body potential cannot be distinguished from a simple pairwise
interaction once it is reduced to its hyperscalar component $V_{0}$ by
suitable angular integration.  It also results from numerical tests
that relativistic kinematics does not change significantly the {\em
relative} magnitude of orbital vs.\ radial excitation energies.

The problem of the Roper resonance is thus a difficult one if one
requires a model to reproduce perfectly the observed masses.  Several
solutions have been proposed, including accepting with philosophy that a
100 MeV discrepancy on a particular level does not necessarily means
the end of the conventional quark level \cite{Isgur:1999jv}.  One
might mention:\\
$1.$ Introducing in the potential some terms with negative Laplacian.
In particular, scalar-meson exchange induces a Yukawa-like potential 
$v\propto -\exp(-\mu r)/r$. Its strength is however not sufficient to 
change the level order, when it is associated with a linear potential 
of reasonable magnitude  \cite{Stassart:2000nk}.\\
$2.$ Introducing a specific spin or spin--flavour dependence in the 
potential, with the consequence that Eq.~(\ref{radial-hyper}) is no 
longer valid: the effective potential in negative-parity states 
differs from the one governing the ground state and its radial 
excitations. This is what occurs with pion exchange in the models 
introduced by Glozman and others.\\
$3.$ More drastic solutions consist of considering the Roper 
resonance as a multiquark state or a hybrid state, possibly mixed 
with conventional radial excitations. The production and decay 
properties do not provide much support for such an interpretation.

Back to spectral theory, the splitting of levels in the nearly
hyperscalar potential (\ref{pot-hyper}) is very similar to the famous
pattern of the  $N=2$ h.o\ multiplet \cite{Gro2,IsgKar}, except that the
Roper is disentangled \cite{Richard:1990ra,JMRrep}.  A similar result
is found for higher negative-parity excitation: the split $N=3$ levels
of the nearly harmonic model are separated into a radially excited
$L=1$ and a set of split $L=3$ levels.  Details and references can be
found in Refs.~\cite{Stancu:1991cz,Richard:1990ra,JMRrep}. 
\section{Test of flavour independence for baryons}
\label{se:baryons:convexity}
A look at the baryon spectrum gives guide lines to our intuition which
are contradictory.  We have just seen that the ``vertical spectrum'', 
i.e., the ordering of radial vs.\
orbital excitations suggests flavour-dependent potentials such as
those mediated by exchange of Goldstone bosons.  On the other hand,
the ``horizontal spectrum'', i.e., the spectrum of ground states with
different flavour content, is rather smooth.  There is a very
continuous and slow increase in mass when a strange quark replaces an
ordinary one.  Strangeness excitation usually costs less than spin
excitation.

This latter observation suggests describing all ground states with the
same universal (flavour independent) potential and then applying moderate
spin corrections ``{\`a} la DGG'' \cite{DGG}, where flavour dependence
occurs only through the mass of the constituent in the Breit--Fermi
term.  This was basically the substance of the Isgur--Karl model
\cite{Isgur:1999jv} and other attempts with non-harmonic confinement
\cite{RTannals}.

This leads to considering the mathematical properties of the  3-body
spectrum in a given potential as a function of the constituent
masses.  We have seen in Sec.~\ref{se:mesons} that for any 2-body
potential $V$ which is flavour independent (just required to produce
bound states), inequalities such as $(b\bar{b})+(c\bar{c})\le
2(b\bar{c})$ can be derived.  The analogue for baryons,
\begin{equation}
    \label{conv-baryon}
    (QQq)+(Q'Q'q)\le 2 (QQ'q),
    \end{equation}
requires, however, mild restrictions, which are always satisfied by the
potentials used in quark model, whatever quark masses are used for
$Q$, $Q'$ and $q$.  Flavour independence is the only serious
hypothesis. See Ref.~\cite{Liebetc}.
    
For instance, the equal spacing rule, $\Omega^- - 
\Xi^*=\Xi^*-\Sigma^*=\Sigma^*-\Delta$, is understood as follows: the 
central force gives a concave behaviour, with for instance  
$\Omega^- - \Xi^*<\Xi^*-\Sigma^*$, but a quasi perfect linearity is 
restored by the spin--spin interaction which acts more strongly on 
light quarks. A similar scenario holds for the Gell-Mann--Okubo 
formula \cite{RTannals}.

Inequalities can also be written for baryons with heavy flavour, some 
of them being more accessible than others to experimental checks in the 
near future. Examples are
\begin{eqnarray}
    \label{charmed-baryons}
   &&3 (bcs)\ge (bbb)+(ccc)+(sss),\\
   &&2 (bcq)\ge (bbq)+(ccq),\qquad 
   2 (cqq)\ge (ccq) + (qqq).\nonumber
\end{eqnarray}
\section{Baryons with two heavy flavours}%
\label{se:two-heavy}
There is a renewed interest in this subject
\cite{Savage,Gershtein:2000nx}.  The recent observation of the
$(b\bar{c})$ mesons demonstrates our ability to reconstruct hadrons with two
heavy quarks from their decay products.

Baryons with two heavy quarks $(QQ'q)$ are rather fascinating: they
combine the adiabatic motion of two heavy quarks as in $J\!/\!\Psi$ and
$\Upsilon$ mesons with the highly relativistic motion of a light quark
as in flavoured mesons $D$ or $B$.

The wave function of $(QQq)$ exhibits a clear diquark clustering with 
$r(QQ)\ll r(Qq)$ for the average distances. This does not necessarily 
mean that for a given potential model, a naive two-step  calculation 
is justified. Here I mean: estimate first the $(QQ)$ mass using the 
direct potential $v(QQ)$ only, and then solve the $[(QQ)-q]$ 2-body 
problem using a point-like diquark. If $v$ is harmonic, one would miss 
a factor $3/2$ in the effective spring constant of the $(QQ)$ system, 
and thus a factor $(3/2)^{1/2}$ in its excitation energy.

On the other hand, it has been checked that the Born--Oppenheimer 
approximation works extremely well for these $(QQq)$ systems, even 
when the quark mass ratio $Q/q$ is not very large. This system is the 
analogue of H$_{2}^+$ in atomic physics.
\section{\boldmath $S_{N}$-expansion \unboldmath and stability of 
multiquarks}
The deuteron and other nuclei are bound by long-range nuclear forces,
not by direct quark--quark interaction,and thus cannot be considered
as genuine multiquark states.  Stable multiquarks have been searched
for in all possible channels, with the exception of very exotic
hadronic configurations with several heavy flavours, and the results
are negative.

The absence, or at least the non-proliferation of stable multiquarks,
is reproduced in most model calculations, even the simplest ones.

I shall not discuss too much here the multiquark states such as the
hexaquark $H$ \cite{JaffeH} or the pentaquark $P$ \cite{Pentaquark}
whose binding is tentatively achieved by chromomagnetic forces.  For
these states, the predictions vary from one author to another, and the
questions are: \\
{\sl i)} Is the one-gluon-exchange appropriate for describing
hyperfine effects ?  Note that the Goldstone-boson-exchange model
gives slightly different predictions on multiquarks \cite{Stancu:1999xc}, as compared to
the chromomagnetic model, but the main message remains identical:
there is no proliferation of multiquarks in the spectrum.  
\\
{\sl ii)}
What is the value of the short-range correlation in a multiquark, as
compared to ordinary hadrons?  This quantity governs, indeed, the
strength of chromomagnetic effects.

Another issue deals with the absence of collective binding due to 
chromoelectric (confining) forces. Consider a four-body problem with 
Hamiltonian
\begin{eqnarray}
    &&H_{4}(x)=\sum_{i=1}^4{\vec{p}_{i}^2\over 2m}+(1-2x)(V_{12}+V_{34})+
    (1+x) (V_{13}+V_{14}+V_{23}+V_{24})\nonumber\\
    &&\phantom{H_{4}(x)}=H_{\rm S}+x H_{\rm MS},
\end{eqnarray}
where the parameter $x$ measures the departure from a fully  
symmetric interaction. From the variational principle, the ground-state energy 
$E(x)$ is maximal at $x=0$. Indeed, with obvious notation, 
\begin{equation}
    E(x)=\langle \Psi(x)\vert H(x)\vert\Psi(x)\rangle\le
    \langle \Psi(0)\vert H(x)\vert\Psi(0)\rangle=E(0),
\end{equation}
where the last step results from $\Psi(0)$ being symmetric and thus
insensitive to the mixed-symmetry part, $H_{\rm MS}$.  As $x$ enters
$H$ linearly, $E(x)$ is a concave function \cite{Thir} and thus should
exhibit near $x=0$ a parabola-like shape.  This
means that
\begin{equation}
    E(0)> E(x_{1})>E(x_{2})\qquad \hbox{if}\qquad 0<x_{1}<x_{2}
\end{equation}
is guaranteed, and
\begin{equation}
   E(x'_{1})>E(x_{2})\qquad \hbox{if}\qquad x'_{1}<0<x_{2}\qquad 
   \hbox{and}\qquad \vert x'_{1}\vert< x_{2}
\end{equation}
is very likely. In short, a larger asymmetry leads to a lower algebraic 
energy.

Consider now the most primitive ansatz for the multiquark interaction, 
the so-called additive colour model
\begin{equation}
    \label{col-pot}
    V=-{3\over 16} \sum_{i<j}
    \tilde{\lambda}_{i}^{\rm c}.\tilde{\lambda}_{j}^{\rm c}\, v(r_{ij})
    =\sum_{i<j} g_{ij}\,v(r_{ij}).
\end{equation}
This is a pairwise interaction with a colour dependence that corresponds
to pure octet exchange.  The normalisation is such that $v$ is the
central potential for quarkonia.  If this model is applied to
four-quark configurations $(\bar{q}\bar{q}qq)$ assuming a frozen colour
wave function, then when one looks at the distribution of the
strength factors $g_{ij}$, one observes that the largest asymmetry is
obtained for the meson-meson threshold, for which an appropriate
labelling gives
\begin{equation}
    g_{ij}=0,\qquad \hbox{except}\qquad g_{12}=g_{34}=1.
\end{equation}
It is easily checked that the strength distributions corresponding to 
a  ``true baryonium'', with $(qq)$ in a $\bar{3}$ colour 
state, and $(\bar{q}\bar{q})$ in $3$ colour state, or a ``mock 
baryonium'' state, with a
$6-\bar{6}$ colour structure, are  more clustered around the  
symmetry point where all $g_{ij}$ are equal to 1/3. (Colour neutrality 
ensures that $\sum g_{ij}$ is the same for all colour-singlet configurations.)

Indeed, solving the equal-mass $(\bar{q}\bar{q}qq)$ problem using a 
frozen colour wave function and the potential (\ref{col-pot}) never leads to 
stable multiquarks.

Binding requires a combination of\\
$-$ mixing of several colour wave functions\\
$-$ spin-dependent effects\\
$-$ a better estimate of the multiquark potential. Note however that the recent 
lattice estimate by Green et al.\ \cite{Green:2000mf} suggests that the true 4-quark potential is 
less attractive than the ansatz (\ref{col-pot}).\\
$-$ another type of $S_{N}$ breaking which would be favourable to 
multiquark binding.

This latter mechanism was suggested several years ago. Its most 
remarkable feature is that theorists agree on the main result, namely 
that $(\bar{q}\bar{q}QQ)$ becomes stable in the limit where the quark 
mass ratio $Q/q$ becomes very large.

The Hamiltonian
\begin{equation}
    H={1\over 2m} (\vec{p}_{1}^2+\vec{p}_{2}^2)
    +{1\over 2M} (\vec{p}_{3}^2+\vec{p}_{4}^2)+V,
\end{equation}
with $V$ even under charge conjugation, i.e., simultaneous 
$(1\leftrightarrow 3)$ and $(2\leftrightarrow 4)$ exchanges, can be 
seen as the $x=1$ value of
\begin{eqnarray}\label{C-decomp}
    &&H(x)=H_{\rm even}+ x H_{\rm odd},\nonumber\\
    &&H_{\rm even}=\left({1\over 4m}+{1\over 4 
    M}\right)\sum_{i}\vec{p}_{i}^2+V,\\
    &&H_{\rm odd}=\left({1\over 4m}-{1\over 4 
        M}\right)\left(\vec{p}_{1}^2+\vec{p}_{2}^2-
       \vec{p}_{3}^2-\vec{p}_{4}^2\right).\nonumber
\end{eqnarray}
When $x$ grows from 0 to 1, the meson--meson threshold is unchanged, 
because the inverse reduced mass of $(m,M)$ is already an average of 
$m^{-1}$ and $M^{-1}$. On the other hand, the four-quark energy 
$E(x)$ decreases. Stability of 
$(\bar{q}\bar{q}QQ)$ becomes possible if $Q\gg q$.

Remarks on this multiquark are in order:\\
$1.$ If the mass ratio approaches the critical value for stability, 
the wave function has large meson--meson components. For larger $Q/q$, 
the binding is more pronounced, and the system acquires another structure 
in space and colour. The two heavy quarks form a colour $\bar{3}$ 
source (as in every $\Xi_{QQ}$ baryon), which in turn, form a colour 
singlet with the two antiquarks, as in every flavoured antibaryon 
$(\bar{Q}qq)$. One thus uses well established colour coupling. This 
contrasts with the highly speculative colour structure in 
``colour-chemistry''.\\
$2.$ The same mechanism holds for $(+,+,-,-)$ configurations in atomic
physics.  The positronium molecule Ps$_{2}$ is very weakly bound.  The
hydrogen molecule, where different masses experience the same
potential as Ps$_{2}$, is much more stable.  In fact the stability of
Ps$_{2}$, as first demonstrated in 1947 by Hylleraas and Ore
\cite{Hylleraas47a}, implies stability for hydrogen and any similar
configurations such as $K^+K^+\pi^-\pi^-$ \cite{Richard93}.  On the
other hand, the molecule $p\bar{p}e^+e^-$ is unstable, even in the
limit where strong interaction and electromagnetic annihilation is
switched off: one can perform an $S_{N}$ expansion analogous to
(\ref{C-decomp}); the 4-body energy decreases as asymmetry
($m^{-1}\neq M^{-1})$ is implemented, but the threshold benefits much
more from this effect.  In the limit where $M\gg m$, the system breaks
in a point-like protonium which cannot attract the remaining
positronium.  The critical mass ratio for stability has been estimated
to about $M/m\simeq 2.2$ \cite{Bressanini97}.
\section{Hall--Post inequalities for mesons and baryons}
The simple additive model (\ref{col-pot}) gives for baryons  the  "1/2" rule 
\begin{equation}
    V_{qqq}=\sum_{i<j}V_{qq}(r_{ij}),\qquad\hbox{with}\qquad
    V_{qq}={1\over2}V_{q\bar{q}},
\end{equation}
or, for Hamiltonians, after distributing the kinetic energy among different 
terms
\begin{equation}
 H_{qqq}={1\over2}\sum_{i<j}H_{q\bar{q}}={1\over2}
 \sum_{i<j}\left[{\vec{p}_{i}^2\over2m}+{\vec{p}_{j}^2\over2m}+
 V_{q\bar{q}}(r_{ij})\right].
\end{equation}
The variational principle, using the baryon wave function gives
\begin{equation} E(qqq)\ge3  E(q\bar{q})/2,\end{equation}
or for the masses, including the contribution of the constituents,
\begin{equation} \label{mes-bar}
   (qqq)/3\ge (q\bar{q})/2.
\end{equation}
In other words, matter is heavier per quark in baryons than in mesons. 
This is seen, e.g., by comparing $\Omega^-(1672)$ and $\phi(1020)$,
where we are dealing with the spin-triplet potential, including the
spin-spin term.  Otherwise, one can compare the spin 1/2 baryon to a
suitable combination of spin 0 and spin 1 mesons.

Some generalisations to unequal quark masses are straightforward, for 
instance
\begin{equation}
   2 (bcs)\ge (b\bar{ c})+(c\bar{ s})+(s\bar{ b})
 \end{equation}
On the other hand, if one rewrites (\ref{mes-bar}) as
\begin{equation}
(\bar{q}\bar{q}\bar{q})+(qqq)\ge 3(\bar{q}q),
\end{equation}
which expresses the fact that quark rearrangement is energetically allowed for 
annihilation at rest, one observes that for a large enough mass ratio 
$Q/ q$, the inequality becomes inverted, namely
\begin{equation}
(\overline{Q}\overline{Q}\overline{Q})+(qqq)< 3(\overline{Q}q).
\end{equation}
For instance, a triple-charm antibaryon would not ``annihilate'' on 
ordinary matter.

These inequalities were rediscovered in quark model studies
\cite{Nus1,AdRiTa,JMR2} but they were known from previous works on the
stability of matter and the thermodynamic limit of $N$-body systems
\cite{JMLL}, and studies of few-nucleon systems \cite{Post}.

The simplest result is the following. If 
\begin{equation}
    H_{N}(m,g)=\sum_{i}^N{\vec{p}_{i}^2\over 2m} +g \sum_{i<j}V(r_{ij})
\end{equation}
governs a system of $N$ bosons and $E_{N}$ denotes the ground-state 
energy, then the identity
\begin{equation}
H_{N}(m,g)= {1 \over N-2}
\sum_{i} H_{N-1}^{(i)}\left({N-1\over N-2}m,g\right), \label{HNH2}
\end{equation}
leads to 
\begin{equation}
E_{N}(m,g) 
\ge {N\over N- 2} E_{N-1}\left({N-1\over 
N-2}m,g\right).
\label{ENE2}
\end{equation}
This inequality is never
saturated, because the overall translation energy of the $(N-1)$-body
subsystems within the $N$-body system is neglected.  An improvement
consists of replacing the decomposition (\ref{HNH2}) by the identity 
\cite{Post,BMR1}
\begin{equation}
\widetilde{H}_{N}(m,g)={1\over
N-2}\sum_{i}\widetilde{H}_{N-1}^{(i)}\left({N\over 
N-1}m,g\right),
\label{HNtildeHN-1}
\end{equation}
relating the translation-invariant Hamiltonians
\begin{equation}
\widetilde{H}_{N}=H_{N}-{[\sum\vec{p}_{i}]^2\over 2 N m}\cdot
\label{HNtilde}
\end{equation}
This leads to the new inequality
\begin{equation}
E_{N}(m,g)
\ge  {N\over N-2} E_{N-1}\left({N\over N-1}m,g\right).
\label{ENtildeE2}
\end{equation}
As $Nm/(N-1)$ is smaller than $(N-1)m/(N-2)$ for $N\ge3$, the bound
(\ref{ENtildeE2}) is better than (\ref{ENE2}), since any binding
energy in a given potential is a decreasing function of the
constituent mass.

By recursion, an inequality such as (\ref{ENE2}) or (\ref{ENtildeE2})
provides a bound on $E_{N}$ in terms of the 2-body energy $E_{2}$.  In
the case of the harmonic oscillator for bosons, the bound derived from
(\ref{ENtildeE2}) is saturated, whereas the bound derived from
(\ref{ENE2}) is, by a factor $\sqrt 2$, smaller than the exact result at
large $N$.

The generalisation to  different constituent masses has been carried 
out \cite{BAMA93}. The decomposition of the Hamiltonian involves some free 
parameters which can be adjusted to optimise the inequality. Again, 
saturation is obtained in the case of harmonic forces.

The situation is far more difficult for fermions. In Ref.~\cite{JMLL}, the 
decomposition
\begin{equation}
H_{N}(m,g)
   ={1\over 2}\sum_{i}\sum _{j\neq i}
 \left[ {\vec{p}_{j}^2\over (N-1) m} + g V(r_{ji})\right]
  \label{HNLL}
\end{equation}
is written down, which expresses $H_{N}$ 
in terms of Hamiltonians with $(N-1)$ independent particles. For fermions, 
one obtains
\begin{equation}
E_{N}(m,g)
   \ge {N\over 2}f_{N-1}(m(N-1),g),
  \label{HNferm}
\end{equation}
where $f_{N}$ is the cumulated energy of a system of $N$ independent
fermions, a notation borrowed from Ref.~\cite{BAMA96}. In 
Eq.~(\ref{HNferm}), the translation energy of the $(N-1)$-body 
subsystem is neglected. 

Basdevant and Martin\cite{BAMA96} succeeded in improving this inequality in the 
case of power-law potentials $V\propto r^n$ with $n>1$ or combinations 
of those with positive weights. They obtained saturation for any 
number $N$ of fermions in the harmonic limit $n=2$. Recently Juillet 
et al.~\cite{Juillet:2000ie} used group-theoretical considerations to analyse the 
structure of $N$-body wave function in terms of the permutation of a 
$(N-1)$-subset and derived the following inequality
\begin{eqnarray}
    \label{EvsN-1F}
\lefteqn{E_N^S(m,g)\ge {N-1\over N(N-2)(2S+1)}\times}\\
&&\left[ S(N+2S+2)
E_{N-1}^{S-1/2}\left(m,{Ng\over N-1}\right)  
+ (S+1)(N-2S)
E_{N-1}^{S+1/2}\left(m,{Ng\over N-1}\right)\right]\cdot\nonumber
\end{eqnarray}
where $S$ is the spin of the $N$-fermion system.

\section{Convergence of variational calculations}
When computing the binding energy of a baryon or a more complicated 
system in a given potential model, one is faced with the choice of a 
method for solving the few-body  problem.

Almost all possible methods have been applied to baryons.  We already
mentioned the approximation of a perturbed oscillator.  The hyperspherical
expansion and the Faddeev equations have also been applied very convincingly. 
The approximation of Feshbach and Rubinow \cite{FeshRub}, which looks at the
best function of the perimeter $\sum r_{ij}$, is not very accurate
for baryons with linear confinement, while it turns out to be very powerful
for weakly-bound states with short-range interaction.

Many variational methods are base on a decomposition
\begin{equation}
    \label{var-exp}
\Psi^M=\sum_{i=1}^{M}c_{i}\varphi_{i}(\alpha,\beta,\ldots),
\end{equation}
where the linear parameters $c_{i}$ are determined from a eigenvalue 
equation and the range parameters $\alpha,\beta,\ldots$ adjusted by 
suitable minimisation of the energy.
 For most choices of the basis functions $\varphi_{i}$, one achieves 
 a decent convergence of the variational energy $E^{N}$ toward the 
 exact energy $E$, with the consequence that the approximate 
 variational wave function $\Psi^N$ overlaps better and better the exact 
  one, i.e., \cite{Thir}
\begin{equation}
    \left\vert \langle\Psi^M\vert\Psi\rangle\right\vert\to 1
    \qquad\hbox{as}\qquad E^M\to E.
\end{equation}
However, the variational wave function $\Psi^M$ might differ
appreciably from the exact $\Psi$ in regions of negligible
contribution to the normalisation integral but of interest for
physics.  In particular, the value of the wave function for 
$r_{ij}\to 0$ governs the production rate, the electromagnetic and
annihilation width of quarkonia, the contribution of $W$-exchange to
the weak decay of flavoured mesons and baryons, the parity-violating
effects in atoms, etc.  A comparison of the estimates of the
correlation coefficient
\begin{equation}
\delta_{12}=\langle\Psi\vert\delta^{(3)}(\vec{r}_{12})\Psi\rangle
\end{equation}
using different 3-body methods is rather interesting. It shows that:\\
$1.$ The quality of the Faddeev wave function is impressive, even with 
a small number of partial waves included.\\
$2.$ The hyperspherical formalism provides a slow but safe 
convergence for this quantity.\\
$3.$ Some variational expansions of type (\ref{var-exp}) fail in 
reproducing $\delta_{12}$ ! This is observed in particular when the 
basis is built out of the successive eigenstates of the harmonic oscillator.
 (These wave functions fall off too rapidly at large separation. To 
 keep the normalisation and the average energy correct, a distortion 
 is introduced at small separation). 
$4.$ However, the expansion on a basis of Gaussians does not suffer 
from the same difficulties, as contributions of longer and longer range can 
always be introduced in the expansion.

\section{Bounds on short-range correlations in baryons}
\label{se:short}
As seen in the last section, it is rather difficult to calculate the 
3-body (or higher) wave function with good zero-range properties.

To avoid lengthy computations, one can estimate $\delta_{12}$ using a
trick proposed by Schwinger for the 2-body problem \cite{QR1} and
generalised to larger systems \cite{HSF,HO2T}.  The basic idea is that
$A$ and $A_{B}=A+[H,B]$ have same expectation value within any eigenstate of
$H$.  A suitable choice of $B$ might make $\langle \Psi^M\vert
A_{B}\vert\Psi^M\rangle$ converge faster than $\langle\Psi^M\vert
A \vert\Psi^M\rangle$ to the exact $\langle\Psi\vert A
\vert\Psi\rangle$ as the size $M$ of the basis is increased.

For a 2-body problem with a central potential $V(r)$, the operator
$A=\delta^{(3)}(\vec{r})$ is replaced by $A_{B}\propto {\rm d}V/{\rm
d}r$, which is much less sensitive to a local flaw of the wave
function.  In particular $\delta_{12}$ comes out exact for a linear
potential, however bad is the trial wave function, provided it is
normalised.  One also understands why in such a purely linear
potential, $\delta_{12}^n$ is the same for all $S$-states, independent
of the radial number $n$.  A better result has been established
\cite{GMbook}: $\delta_{12}^n$ increases with $n$ if $V''>0$, and
decreases if $V''<0$.  For instance, current potentials $-a/r+br$
explain why the leptonic coupling is smaller for radially excited
quarkonia than for the ground state.

For baryons, $A_{B}$ contains also a small centrifugal term, besides 
the derivative of the full potential $V$ with respect to $r_{12}$. 
Cohen and Lipkin \cite{CohLip} have compared the $\Sigma$ and $\Xi$ 
hyperfine splittings in models where this is induced by a contact term.
In other words, they compared $\delta_{12}$ in $(qqs)$ and $(ssq)$. 
On can also \cite{JMRshortrange} derive bounds on $\delta_{12}$. 

The short-range correlation between electrons in 
ordinary atoms, the annihilation probability of exotic molecules with 
positrons can also be calculated using the Schwinger rule.

\section*{Acknowledgments}
I would like to thank very much the organizers for the stimulating 
atmosphere of  this Conference, and John Cole for constructive comments.
%

\end{document}